\documentclass[twocolumn,aps,prl,showpacs]{revtex4}

\usepackage{latexsym,amssymb,float}
\usepackage{setspace}
\usepackage{graphicx}
\usepackage{epsfig}

\newcommand{\e}{\varepsilon}

\newcommand{\f}{\frac}
\newcommand{\up}{\uparrow}
\newcommand{\down}{\downarrow}

\begin{document}

%\title{The resonance peak in the electron-doped cuprate superconductors}
\title{Magnetic Resonance in the Spin Excitation Spectrum of Electron-Doped Cuprate Superconductors}
\author{J.-P. Ismer$^{1,2}$, Ilya Eremin$^{1,2}$, Enrico Rossi$^{3}$,  Dirk K. Morr$^{3}$}

\affiliation{$^1$ Max-Planck
Institut f\"ur Physik komplexer Systeme, D-01187 Dresden, Germany \\
$^2$ Institute f\"ur Mathematische und Theoretische Physik,
Technische Universit\"at Carolo-Wilhelmina zu Braunschweig, 38106
Braunschweig, Germany \\
$^3$ Department of Physics, University of Illinois at Chicago,
Chicago, IL 60607}
\date{\today}

\begin{abstract}
We study the emergence of a magnetic resonance in the
superconducting state of the electron-doped cuprate superconductors.
We show that the recently observed resonance peak in the
electron-doped superconductor
Pr$_{0.88}$LaCe$_{0.12}$CuO$_{4-\delta}$ is consistent with an
overdamped spin exciton located near the particle-hole continuum. We
present predictions for the magnetic-field dependence of the
resonance mode as well as its temperature evolution in those parts
of the phase diagram where $d_{x^2-y^2}$-wave superconductivity may
coexist with an antiferromagnetic spin-density wave.
\end{abstract}

 \maketitle

Recently, inelastic neutron scattering (INS) experiments on the
electron-doped high-temperature superconductors (HTSC)
Pr$_{0.88}$LaCe$_{0.12}$CuO$_{4-\delta}$ (PLC$_{0.12}$CO)
\cite{dai} observed a resonance peak in the superconducting (SC)
state, a phenomenon similar to that observed in the hole-doped
cuprates \cite{rossat,exp,Fong95}. While the resonance frequency
in PLC$_{0.12}$CO, $\omega_{\mathrm{res}} \approx 11$ meV, obeys
the same scaling with $T_c$ as that in the hole-doped HTSC, there
exist two significant differences. First, the resonance is
confined to a small momentum region around ${\bf Q}=(\pi,\pi)$,
where it is almost dispersionless. Second, angle-resolved
photoemission (ARPES) experiments on PLC$_{0.11}$CO
\cite{takahashi} estimated, based on measurements of the leading
edge gap, a maximum SC gap located at the ``hot spots" [the Fermi
surface (FS) points connected by ${\bf Q}$] of
$\Delta_{\mathrm{hs}} \approx 5$ meV. Assuming the same SC gap in
PLC$_{0.12}$CO, this would suggest that the resonance is located
slightly above the onset of the particle-hole (\emph{p-h})
continuum given by $2 \Delta_{\mathrm{hs}}$, a result which would
challenge the interpretation of the resonance as a spin exciton
\cite{Fong95,liu,norman2,eremin1,Li}. However, the uncertainties
in ARPES and INS experiments are currently such that it is not
possible to determine the relative magnitude of
$\omega_{\mathrm{res}}$ and $2 \Delta_{\mathrm{hs}}$ and thus
ascertain the validity of the spin exciton scenario.

In this Letter we address this issue and study the emergence of a
resonance mode in the SC state of electron-doped HTSC. We show
that the experimental features of the resonance in PLC$_{0.12}$CO
can be explained within a spin-exciton scenario. In particular, we
demonstrate that the position of the hot spots close to the
Brillouin zone (BZ) diagonal \cite{onufrieva,krotkov} combined
with the momentum dependence of the fermionic interaction leads to
an almost dispersionless resonance that is confined to a small
momentum region around ${\bf Q}$. Moreover, while the resonance is
always located below the \emph{p-h} continuum in systems with a
quasiparticle lifetime, $1/\Gamma \rightarrow \infty$, we show
that the maximum of the resonance's intensity can be shifted to
frequencies above the \emph{p-h} continuum when $1/\Gamma$ is
sufficiently small. In this case, the form of the spin
susceptibility is more reminiscent of the {\it magnetic coherence
effect} in La$_{2-x}$Sr$_x$CuO$_4$ \cite{mason,morr} than of the
resonance observed in the hole-doped HTSC. We present two
predictions for further experimental tests of the spin exciton
scenario. First, we show that a magnetic field in the $ab$ plane
leads to an energy splitting of the resonance which for typical
fields is sufficiently large to be experimentally observable in
the electron-doped HTSC. Second, we predict that in those parts of
the phase diagram, where $d_{x^2-y^2}$-wave superconductivity
(\emph{d}SC) coexists with an antiferromagnetic spin-density wave
(SDW) \cite{AForder1,AForder2,daiprep} and $T_N < T_c$, the
resonance evolves into the Goldstone mode of the SDW state as
$T_N$ is approached.

\begin{figure}[!h]
\epsfig{file=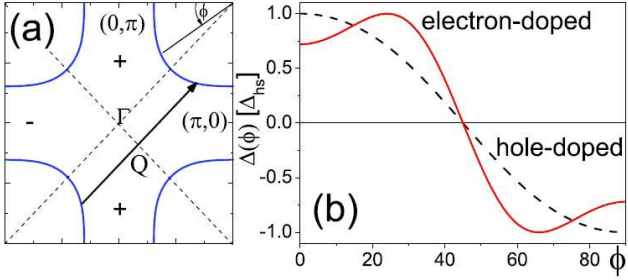,width=8.5cm}
 \caption{(color online). (a) Fermi surface in optimally
electron-doped cuprates. The arrow indicates the scattering of
quasiparticles by {\bf Q}. (b) Angular dependence of the SC gap for
electron and hole-doped HTSC.} \label{fig1}
\end{figure}

The starting point for our study of the resonance mode in the
electron-doped cuprates is the Hamiltonian
\begin{equation}
H = \sum_{{\bf k} \sigma} \varepsilon_{\bf k} c_{{\bf
k},\sigma}^{\dagger} c_{{\bf k}, \sigma}  + \sum_{{\bf k}}
\Delta_{\bf k} c_{{\bf k},\uparrow}^{\dagger} c_{{\bf
-k},\downarrow}^{\dagger} + \mathrm{H.c.}  , \label{hubbard}
\end{equation}
where $c_{{\bf k},\sigma}^{\dagger}$ creates an electron with spin
$\sigma$ and momentum ${\bf k}$, and $\Delta_{\bf k}$ is the SC gap
with $d_{x^2-y^2}$-wave symmetry. The normal state tight binding
dispersion
\begin{eqnarray}
\epsilon_{\bf k} & = & -2t \left( \cos k_x + \cos k_y \right) - 4
t'
\cos k_x \cos k_y  \nonumber \\
&& - 2t'' \left( \cos 2k_x + \cos 2k_y \right) - \mu
\end{eqnarray}
with $t=250$ meV, $t'/t=-0.4$, $t''/t=0.1$ and $\mu/t=-0.2$
reproduces the position of the hot spots and the underlying FS
[Fig.~\ref{fig1}(a)] as inferred from ARPES \cite{takahashi}.

Despite the same FS topology in the electron-doped and hole-doped
cuprates, the angular dependence of the superconducting gap along
the FS is qualitatively different in these systems. Based on a
scenario in which superconductivity arises from the exchange of
antiferromagnetic spin fluctuations, it was argued that the
maximum SC gap is achieved near the hot spots
\cite{krotkov,blumberg,manske1}. In the hole-doped cuprates, the
hot spots are located close to ${\bf q}=(\pm \pi,0)$ and $(0,\pm
\pi)$, resulting in a SC $d_{x^2 - y^2}$-wave gap that varies
monotonically along the FS, as shown in Fig.~\ref{fig1}(b). In
contrast, in the electron-doped cuprates, the hot spots are
located much closer to the zone diagonal [Fig.~\ref{fig1}(a)],
leading to a nonmonotonic behavior of the SC gap
\cite{krotkov,manske1}, in agreement with ARPES experiments
\cite{takahashi}. A good fit of $\Delta_{\bf k}$ to the
experimental data is achieved via the inclusion of a higher
harmonic, such that $\Delta_{\bf k} = \frac{\Delta_0}{2} \left(
\cos k_x - \cos k_y \right)  + \frac{\Delta_1}{2} \left( \cos 2k_x
- \cos 2k_y \right)$, where $\Delta_1 /\Delta_0 = 0.63$ ensures
that the maximum of $|\Delta_{\bf k}|$ along the FS is located at
the hot spots, as shown in Fig.~\ref{fig1}(b) \cite{Schnyder}.
Since the magnitude of the SC gap in PLC$_{0.12}$CO is still
unknown, we use $\Delta_0= 10$ meV which yields
$\Delta_{\mathrm{hs}}=5$ meV thus reproducing the ARPES estimate
of the SC gap at the hot spots of PLC$_{0.11}$CO. Note that the
results shown below are robust against (reasonable) changes in the
band structure, as long as $\Delta_{\mathrm{hs}}$ remains
unchanged.
%
% Fig.1
%

Similar to the hole-doped HTSC \cite{Fong95,liu,norman2,eremin1} the
resonance peak in the SC state of the electron-doped cuprates can be
understood by considering the dynamical spin susceptibility within
the random phase approximation (RPA)
\begin{equation}
\chi({\bf q},\omega)=\frac{\chi_0({\bf q},\omega)}{1-U({\bf q})
\chi_0({\bf q},\omega)} , \label{fullchi}
\end{equation}
where $U({\bf q})$ is the fermionic four-point vertex and
$\chi_0({\bf q},\omega)$ is the free-fermion susceptibility given
by the sum of two single bubble diagrams consisting of either
normal or anomalous Green functions. While momentum independent as
well as momentum dependent forms of $U({\bf q})$ were used in the
hole-doped cuprates \cite{Fong95,liu,norman2,eremin1}, the close
proximity of the (commensurate) antiferromagnetic and SC phases in
the electron-doped HTSC suggests that $U({\bf q})$ is momentum
dependent, with a maximum at ${\bf Q}=(\pi,\pi)$ \cite{greven}.
Here, we use $U_{\bf q}=-\f{U_{0}}{2}(\cos q_{x} + \cos q_{y})$
which reproduces a nearly dispersionless resonance mode around
${\bf Q}$. The form of $\chi_0$ in the hole-doped and
electron-doped HTSC is qualitatively similar, and has been
extensively discussed for the former
\cite{Fong95,liu,norman2,eremin1}. For momenta ${\bf q}$ near
${\bf Q}$ and $\Gamma = 0^+$, Im$\chi_0$ is zero at low
frequencies and exhibits a discontinuous jump at the onset
frequency of the \emph{p-h} continuum $\Omega_{c}({\bf
q})=|\Delta_{\bf k}|+ |\Delta_{\bf k+q}|$, where both ${\bf k}$
and ${\bf k+q}$ lie on the FS [for ${\bf q}={\bf Q}$ one has
$\Omega_{c}({\bf Q})=2\Delta_{\mathrm{hs}}$]. The discontinuity in
Im$\chi_0$, which is a direct consequence of ${\rm
sgn}(\Delta_{\bf k})=-{\rm sgn}(\Delta_{\bf k+q})$ and hence the
$d_{x^2-y^2}$-wave symmetry of the SC gap, leads to a logarithmic
singularity in Re$\chi_0$. As a result, the resonance conditions
(i) $U_{\bf Q} \mbox{Re}\chi_0({\bf Q},\omega_{\mathrm{res}})=1$
and (ii) Im$\chi_0({\bf Q},\omega_{\mathrm{res}})=0$ can be
fulfilled simultaneously at $\omega_{\mathrm{res}}<\Omega_c$ for
any $U_{\bf Q}>0$, leading to the emergence of a resonance peak as
a spin exciton. Note that for finite $\Gamma$, condition (i) can
only be satisfied if $U_{\bf Q}$ exceeds a critical value, and
condition (ii) is replaced by $U_{\bf Q}{\rm Im}\chi_0({\bf
Q},\omega_{\mathrm{res}})\ll 1$.

%
% Fig.2
%
\begin{figure}[!h]
\epsfig{file=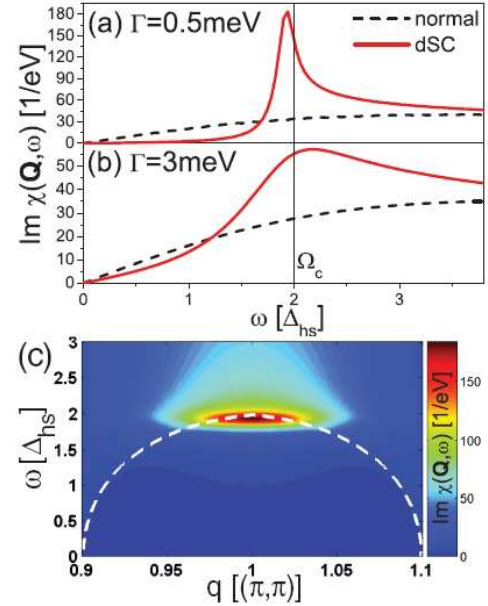,width=6.5cm} \caption{(color online).
Im$\chi({\bf Q },\omega)$ for $U_{0}=0.854$ eV and (a)
$\Gamma=0.5$ meV, and (b) $\Gamma=3$ meV. (c) Contour plot of
Im$\chi$ in the $(\omega,{\bf q})$ plane for the parameters in
(a), together with $\Omega_c({\bf q})$ (dashed white line).}
\label{fig2}
\end{figure}
In Fig.~\ref{fig2}, we present Im$\chi({\bf Q},\omega)$ resulting
from Eq.~(\ref{fullchi}). Since the INS data suggest that the
resonance is located close to the \emph{p-h} continuum, we have
chosen $U_{0}=0.854$ eV such that for $\Gamma \rightarrow 0$ (not
shown), the resonance is located at $\omega_{\mathrm{res}} = 9.8
\enspace {\rm meV}=0.98\Omega_c$. For small $\Gamma=0.5$ meV
[Fig.~\ref{fig2}(a)], the resonance broadens but only exhibits a
negligible frequency shift. However, for larger $\Gamma=3$ meV
[Fig.~\ref{fig2}(b)], the resonance has not only become much
broader but its peak intensity has also shifted to
$\omega_{\mathrm{res}} \approx 11 \enspace {\rm meV}=1.1\Omega_c$
well above the onset of the \emph{p-h} continuum. In this case,
neither of the resonance conditions is satisfied, and Im$\chi$ is
more reminiscent of the magnetic coherence effect in
La$_{2-x}$Sr$_x$CuO$_4$ \cite{mason,morr} than of the resonance in
the hole-doped HTSC. Hence, when the resonance is located close to
the \emph{p-h} continuum, its form depends rather sensitively on
$\Gamma$.

The observed spectral weight in Im$\chi$ at frequencies much below
$\Omega_c$ \cite{dai} is consistent with a shorter lifetime
$1/\Gamma$, arising, for example, from disorder effects. Whether
the INS data are better described by the results shown in
Fig.~\ref{fig2}(a) (albeit with a larger $\Delta_{\mathrm{hs}}$
such that the experimentally determined $\omega_{\mathrm{res}}=11$
meV corresponds to $0.98\Omega_c$), or in Fig.~\ref{fig2}(b), is
presently unclear, mainly due to experimental resolution effects
which are in general difficult to account for. Moreover, since the
resonance's maximum intensity is affected by its distance to the
\emph{p-h} continuum, the band structure, the magnitude of the SC
gap, and $1/\Gamma$, we expect it to be smaller in the
electron-doped HTSC than in the hole-doped cuprates.

In Fig.~\ref{fig2}(c) we present a contour plot of Im$\chi$ for
$\Gamma=0.5$ meV along ${\bf q}=\eta(\pi,\pi)$ together with the
momentum dependence of $\Omega_{c}({\bf q})$, the onset of the
\emph{p-h} continuum. The resonance is almost dispersionless and
exists only in a small momentum region ($0.96 {\bf Q} \lesssim
{\bf q} \lesssim 1.04 {\bf Q}$) around ${\bf Q}$, in agreement
with experiment \cite{dai}. This effect arises from the momentum
dependence of $U({\bf q})$, combined with the fact that the
resonance at ${\bf Q}$ is located only slightly below the
\emph{p-h} continuum, which leads to a ``merging" of the resonance
with the \emph{p-h} continuum at small deviations from ${\bf Q}$.
The momentum connecting the nodal points, $\mathbf{q}_n \approx
0.9 {\bf Q}$, where $\Omega_c({\bf q})$ reaches zero, is much
closer to ${\bf Q}$ than in the hole-doped systems where
$\mathbf{q}_n \approx 0.8 {\bf Q}$, leading to an additional
narrowing of the dispersion.

\begin{figure}[h]
\epsfig{file=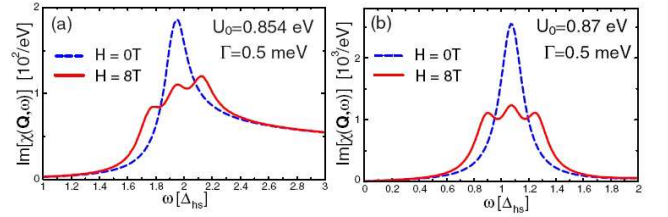,width=8.5cm}
 \caption{(color online).
Im$\chi({\bf Q},\omega)$ for $H=0$ T (dashed blue curve) and $H=8$
T (solid red curve) at $T=2$ K.} \label{fig3}
\end{figure}

The phase diagram of the electron-doped HTSC, and a SC gap that is
much smaller than in the holed-doped cuprates, provide further
opportunities for testing the nature of the resonance peak.
Consider, for example, the effects of a magnetic field $H$ in the
$ab$ plane, which enters the calculation of $\chi$ only through
the Zeeman splitting of the electronic bands, while orbital
effects are absent \cite{norman1}. The field lifts the degeneracy
of the transverse and longitudinal components of $\chi_0$ which
are given by
\begin{eqnarray}\label{Imchi}
\chi_{0}^{\pm \mp} ({\bf q}, i\omega_{n}) &=& -\f{1}{N}\sum_{\bf
k}\left\{c^{+}\f{f^\pm_{\bf k+q}-f^\mp_{\bf
k}}{i\omega_{n}+\xi^\pm_{\bf k+q}-\xi^\mp_{\bf k} }
\right.\nonumber\\&&\left.+\f{c^{-}}{2} \f{1-f^\mp_{\bf
k+q}-f_{\bf k}^\mp}{i\omega_{n}-\xi^\mp_{\bf k+q}-\xi^\mp_{\bf k}}
\right.\nonumber\\&&\left.-\f{c^{-}}{2}
\f{1-f_{\bf k+q}^\pm -f_{\bf k}^\pm}{i\omega_{n}+\xi^\pm_{\bf k+q}+\xi^\pm_{\bf k}}\right\},\nonumber\\
\chi_{0}^{zz} ({\bf q}, i\omega_{n}) &=&-\f{1}{4N} \sum_{\bf
k}\left\{c^{+}\f{f_{\bf k+q}^{+}-f_{\bf k}^{+}}{i\omega_{n}+E_{\bf
k+q}-E_{\bf k}}\right.\\&&\left. +c^{+}\f{f_{\bf k+q}^{-}-f_{\bf
k}^{-}}{i\omega_{n}+E_{\bf k+q}-E_{\bf k}}
\right.\nonumber\\&&\left.+c^{-}\f{1-f_{\bf k+q}^- -f_{\bf
k}^+}{i\omega_{n}-E_{\bf k+q}-E_{\bf
k}}\right.\nonumber\\&&\left.-c^{-}\f{1-f_{\bf k+q}^+ -f_{\bf k}^-
}{i\omega_{n}+E_{\bf k+q}+E_{\bf k}}\nonumber\right\},
\end{eqnarray}
where
$\chi_{0}^{zz}=(\chi_{0}^{\up\up}+\chi_{0}^{\down\down}+\chi_{0}^{\up\down}+\chi_{0}^{\down\up})/4$,
$\xi_{\bf k}^\pm=E_{\bf k}\pm H$, $f_{\bf k}^\pm=f(\xi_{\bf
k}^\pm)$ is the Fermi-function, $c^{\pm} =
\f{1}{2}\left(1\pm\f{\e_{\bf k}\e_{\bf k+q}+\Delta_{\bf
k}\Delta_{\bf k+q}}{E_{\bf k}E_{\bf k+q}}\right)$, and  $E_{\bf
k}= \sqrt{\e_{\bf k}^{2}+\Delta_{\bf k}^{2}}$ (we set
$g\mu_{B}S=1$). The RPA result for the full susceptibility is
\begin{eqnarray}\label{RPAtotal}
\chi&=&\f{1}{2}\left(\f{\chi_{0}^{\pm}}{1-U_{\bf
q}\chi_{0}^{\pm}}+ \f{\chi_{0}^{\mp}}{1-U_{\bf
q}\chi_{0}^{\mp}}\right)
\nonumber\\&&+\f{\chi_{0}^{zz}+\f{1}{2}U_{\bf
q}(\chi_{0}^{\up\up}\chi_{0}^{\down\down}-\chi_{0}^{\up\down}\chi_{0}^{\down\up})}{(1-U_{\bf
q}\chi_{0}^{\up\down})(1-U_{\bf q}\chi_{0}^{\down\up})-U_{\bf
q}^{2}\chi_{0}^{\up\up}\chi_{0}^{\down\down}} .
\end{eqnarray}
The Zeeman term in the denominators of $\chi_0$ shifts
$\Omega_c({\bf q})$ in $\chi_{0}^{\mp}$ ($\chi_{0}^{\pm}$) by $+2H
\approx 1$ meV ($-2H$), while $\Omega_c({\bf q})$ remains
unaffected in $\chi_{0}^{zz}$ \cite{com6}. This effect leads to a
splitting of the resonance into three peaks \cite{Jia06}, as shown
in Fig.~\ref{fig3} for $H=8 {\rm T} \ll H^{ab}_{c2}(T=0) \gtrsim
50$ T \cite{Hc2} [for the results in Fig.~\ref{fig2}(b), the
magnetic field effects are negligible].
%
%Fig3
%
If the resonance is located close to the \emph{p-h} continuum
[Fig.~\ref{fig3}(a)], the splitting results in an asymmetric
resonance, while for a resonance located well below $\Omega_c$,
Im$\chi$ exhibits a symmetric splitting [Fig.~\ref{fig3}(b)].
Hence, the resonance's asymmetry is an indication for the
proximity of the \emph{p-h} continuum. The current experimental
resolution at a frequency of 10 meV is about 1 meV \cite{Daipc},
such that the predicted splitting of the resonance should be
experimentally observable, in contrast to the hole-doped HTSC.

\begin{figure}[!h]
\epsfig{file=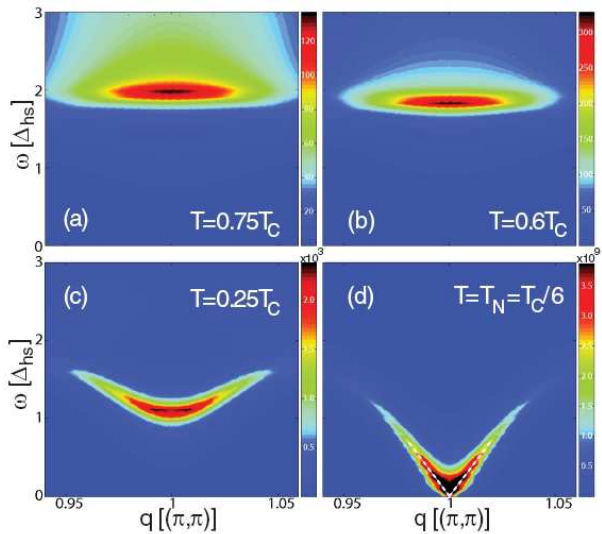,width=8.0cm} \caption{(color online).
Contour plots of Im$\chi$ (in units of 1/eV) for $\Gamma=0.5$ meV
at four different temperatures and $U_0(T)=0.8784 eV - 0.0014eV/K
\cdot T$. The white dashed line in (d) represents the dispersion
of the Goldstone mode. } \label{fig4}
\end{figure}

One of the most important questions regarding the nature of the
resonance mode in the hole-doped cuprates is whether with
decreasing doping, the resonance mode transforms into the
Goldstone mode of the antiferromagnetic parent compounds.
Important insight into this question can be provided by those
electron-doped cuprates, in which superconductivity coexists with
a commensurate SDW \cite{AForder1,AForder2,daiprep}. Extending the
spin-exciton scenario to such a coexistence phase
\cite{schrieffer}, we find that the spin response at $(\pi,\pi)$
only possesses a Goldstone mode at zero energy but no additional
resonance at higher energies. Since the existence of a Goldstone
mode requires the condition $U_{\bf Q}\mbox{Re} \chi_0 ({\bf Q},
\omega)=1$ to be satisfied at $\omega=0$, and since Re$\chi_0
({\bf Q}, \omega )$ increases monotonically with frequency from
$\omega = 0$ up to $\Omega_c$, the resonance conditions can only
be satisfied once, namely at $\omega_{\mathrm{res}}=0$. In
contrast, in a ``pure" SC state, the resonance is necessarily
located at $\omega_{\mathrm{res}} \not = 0$. Moreover, the results
of recent experiments \cite{daiprep,Daipc} suggest the existence
of electron-doped HTSC with $T_N<T_c$. For these materials, it
follows from the above discussion that the resonance of the pure
SC state shifts downward in energy with decreasing temperature,
until it reaches zero energy at $T_N$ and forms the Goldstone
mode. Within the spin exciton scenario, this requires that $U({\bf
q})$ increases with decreasing temperature. In order to exemplify
the resonance's temperature evolution, we consider a system with
$T_N < T_c$ (for concreteness we chose $T_N=T_c/6$). We assume
(somewhat arbitrarily) that $U_0$ varies linearly with temperature
while the momentum dependence of $U({\bf q})$ remains unchanged,
such that the resonance at ${\bf Q}$ and $T=0.75T_c$ is located at
$\omega_{\mathrm{res}} \approx 0.92 \Omega_c$, while for $T=T_N$,
we have $\omega_{\mathrm{res}}=0$. In Fig.~\ref{fig4}, we present
the contour plots of Im$\chi$ for four different temperatures.
%
%Fig4
%
While the resonance shifts downwards with decreasing temperature,
it remains nearly dispersionless over some temperature range below
$T_c$ [see Figs.~\ref{fig4}(a) and \ref{fig4}(b)]. However, upon
decreasing temperature even further [Fig.~\ref{fig4}(c)], the
``flat" dispersion evolves continuously into an upward dispersion
whose minimum is located at ${\bf Q}$. Finally, at $T_N$
[Fig.~\ref{fig4}(d)], the minimum of the dispersion at ${\bf Q}$
reaches zero energy and the resonance becomes the Goldstone mode
of the SDW state \cite{com7}. The qualitative features of this
evolution are independent of the specific temperature dependence
of $U_0$, as long as $U({\bf q})$ decreases sufficiently fast with
deviation from $(\pi,\pi)$.

In summary, we studied the emergence of a magnetic resonance mode
in the SC state of the electron-doped HTSC. We show that the
recently observed resonance peak in PLC$_{0.12}$CO is likely an
overdamped spin exciton located near the \emph{p-h} continuum. We
discuss the magnetic-field dependence of the resonance as well as
its temperature evolution in those parts of the phase diagram
where \emph{d}SC may coexist with a antiferromagnetic SDW.

We would like to thank A.~Chubukov, F.~Kr\"uger, P.~Dai and
J.~Zaanen for helpful discussions and P.~Dai and R.~L.~Greene for
sharing their experimental data prior to publication. This work
was supported by the DAAD under Grant No.~D/05/50420, the
Alexander von Humboldt Foundation, the NSF under Grant
No.~DMR-0513415, the U.S.~DOE under Award No. DE-FG02-05ER46225,
and the IMPRS on "Dynamic Processes in Atoms, Molecules, and
Solids".


\begin{thebibliography}{99}

\bibitem{dai} S.D. Wilson {\it et al.}, Nature (London) {\bf 442}, 59 (2006).

\bibitem{rossat} J. Rossat-Mignod {\it et al.}, Physica (Amsterdam) {\bf 185-189C}, 86 (1991).

\bibitem{exp} H.A. Mook, {\it et al.}, Phys. Rev. Lett. {\bf 70}, 3490
(1993); P. Bourges {\it et al.}, Phys. Rev. B {\bf 53}, 876
(1996); P. Dai {\it et al.}, Phys. Rev. Lett. {\bf 77}, 5425
(1996); H.F. Fong {\it et al.}, Nature (London) {\bf 398}, 588
(1999); P. Dai {\it et al.}, Science {\bf 284}, 1344 (1999); H. He
{\it et al.}, Science {\bf 295}, 1045 (2002); S. Pailh\`{e}s {\it
et al.}, Phys. Rev. Lett. {\bf 93}, 167001 (2004); S.M. Hayden
{\it et al.}, Nature (London) {\bf 429}, 531 (2004); C. Stock {\it
et al.}, Phys. Rev. B {\bf 69}, 014502 (2004).

\bibitem{Fong95}H.F. Fong {\it et al.}, Phys. Rev. Lett. {\bf 75}, 316 (1995).

\bibitem{takahashi} H. Matsui {\it et al.,} Phys. Rev. Lett. {\bf 95}, 017003 (2005).

\bibitem{liu} D.Z. Liu, Y. Zha, and K. Levin,
Phys. Rev. Lett. {\bf 75}, 4130 (1995); A.J. Millis and H. Monien,
Phys. Rev. B {\bf 54}, 16172 (1996); A. Abanov and A.V. Chubukov,
Phys. Rev. Lett. {\bf 83}, 1652 (1999); J. Brinckmann and P.A.
Lee, Phys. Rev. Lett. {\bf 82}, 2915 (1999); I. Sega, P. Prelov\v
sek, and J. Bonca, Phys. Rev. B {\bf 68}, 054524 (2003).

\bibitem{norman2}  M.R.
Norman, Phys. Rev. B {\bf 61}, 14751 (2000); {\bf 63}, 092509
(2001)

\bibitem{eremin1} I.Eremin {\it et al.}, Phys. Rev. Lett. {\bf 94}, 147001 (2005).

\bibitem{Li} J.-X. Li, J. Zhang, and J. Luo, Phys. Rev. B {\bf
68}, 224503 (2003).

\bibitem{onufrieva} F. Onufrieva and P. Pfeuty, Phys. Rev. Lett. {\bf
92}, 247003 (2004).

\bibitem{krotkov} P. Krotkov and A.V. Chubukov, Phys. Rev. Lett. {\bf 96}, 107002
(2006); Phys. Rev. B {\bf 74}, 014509 (2006).

\bibitem{mason} T.E. Mason {\it et al.,} Phys. Rev. Lett. {\bf 77}, 1604
(1996); B. Lake {\it et al.,} Nature (London) {\bf 400}, 43
(1999).

\bibitem{morr} D.K. Morr and D. Pines, Phys. Rev. B {\bf 61}, R6483
(2000); {\bf 62}, 15177 (2000).

\bibitem{AForder1} G.M. Luke {\it et al.,} Phys. Rev. B {\bf 42}, 7981 (1990).

\bibitem{AForder2} P.K. Mang {\it et al.},
Phys. Rev. Lett. {\bf 93}, 027002 (2004).

\bibitem{daiprep} S.D. Wilson {\it et al.,} Phys. Rev. B {\bf 74}, 144514
(2006).

\bibitem{blumberg} G. Blumberg {\it et al.,} Phys. Rev. Lett. {\bf 88}, 107002 (2002).

\bibitem{manske1} D. Manske, I. Eremin, and K.-H. Bennemann, Phys.
Rev. B {\bf 62}, 13922 (2000).

\bibitem{Schnyder} Similar results are also obtained for other higher harmonic terms
in $\Delta_{\bf k}$, see also A.P. Schnyder {\it et al.}, Phys.
Rev. B {\bf 70}, 214511 (2004).

\bibitem{greven} E.M. Motoyama {\it et al.}, Nature (London) {\bf 445}, 186
(2007).

\bibitem{norman1} For orbital effects, see M. Eschrig, M.R. Norman, and B.
Janko, Phys. Rev. B {\bf 64}, 134509 (2001).

\bibitem{com6} The effect of $H$ on the Fermi-functions is
negligible around the hot spots, at least for $T \ll T_c$.

\bibitem{Jia06} Note that in H.-M. Jiang and J.-X. Li, Phys. Rev. B {\bf 73}, 224507
(2006) no splitting of the resonance was found since only $\chi^\pm$
was considered.

\bibitem{Hc2} P. Li, F. Balakirev, and R.L. Greene, Phys. Rev. B {\bf 75}, 172508 (2007).

\bibitem{Daipc} P. Dai (private communication).

\bibitem{schrieffer} J.R. Schrieffer, X.G. Wen, and S.C. Zhang,
Phys.  Rev. B {\bf 39}, 11663 (1989).

\bibitem{com7} Note that for a momentum independent $U({\bf q})$, the emerging SDW
state is incommensurate.

\end{thebibliography}
\end{document}